\documentclass{emulateapj}

\usepackage{natbib}
\usepackage{amsmath}
\usepackage{graphicx}

\shorttitle{The HERA Dish I: Beam Measurements and Science Implications}
\shortauthors{Neben et al.}

\begin{document}

\title{The Hydrogen Epoch of Reionization Array Dish I: Beam Pattern Measurements and Science Implications}


\author{Abraham R. Neben\altaffilmark{1,2},
Richard F. Bradley\altaffilmark{3,4},
Jacqueline N. Hewitt\altaffilmark{1,2},
David R. DeBoer\altaffilmark{5},
Aaron R. Parsons\altaffilmark{5},
James E. Aguirre\altaffilmark{7},
Zaki S. Ali\altaffilmark{5},
Carina Cheng\altaffilmark{5},
Aaron Ewall-Wice\altaffilmark{1,2},
Nipanjana Patra\altaffilmark{5},
Nithyanandan Thyagarajan\altaffilmark{6},
Judd Bowman\altaffilmark{6},
Roger Dickenson\altaffilmark{4},
Joshua S. Dillon\altaffilmark{5},
Phillip Doolittle\altaffilmark{4},
Dennis Egan\altaffilmark{4},
Mike Hedrick\altaffilmark{4},
Daniel C. Jacobs\altaffilmark{6},
Saul A. Kohn\altaffilmark{7},
Patricia J. Klima\altaffilmark{4},
Kavilan Moodley\altaffilmark{8},
Benjamin R.B. Saliwanchik\altaffilmark{8},
Patrick Schaffner\altaffilmark{4},
John Shelton\altaffilmark{4},
H.A. Taylor\altaffilmark{4},
Rusty Taylor\altaffilmark{4},
Max Tegmark\altaffilmark{1,2},
Butch Wirt\altaffilmark{4},
Haoxuan Zheng\altaffilmark{1,2}}

\affil{\altaffilmark{1}MIT Kavli Institute, Massachusetts Institute of Technology, Cambridge, MA, 02139 USA}
\affil{\altaffilmark{2}Dept. of Physics, Massachusetts Institute of Technology, Cambridge, MA, 02139 USA}
\affil{\altaffilmark{3}Dept. of Electrical and Computer Engineering, University of Virginia, Charlottesville, VA 22904}
\affil{\altaffilmark{4}National Radio Astronomy Obs., Charlottesville, VA}
\affil{\altaffilmark{5}Dept. of Astronomy, University of California, Berkeley, CA, USA}
\affil{\altaffilmark{6}Arizona State University, School of Earth and Space Exploration, Tempe, AZ 85287, USA}
\affil{\altaffilmark{7}Dept. of Physics and Astronomy, University of Pennsylvania, Philadelphia, PA}
\affil{\altaffilmark{8}Astrophysics and Cosmology Research Unit, University of KwaZulu-Natal, Durban, South Africa}



\begin{abstract}
The Hydrogen Epoch of Reionization Array (HERA) is a radio interferometer aiming to detect the power spectrum of 21\,cm fluctuations from neutral hydrogen from the Epoch of Reionization (EOR). Drawing on lessons from the Murchison Widefield Array (MWA) and the Precision Array for Probing the Epoch of Reionization (PAPER), HERA is a hexagonal array of large (14\,m diameter) dishes with suspended dipole feeds. Not only does the dish determine overall sensitivity, it affects the observed frequency structure of foregrounds in the interferometer. This is the first of a series of four papers characterizing the frequency and angular response of the dish with simulations and measurements. We focus in this paper on the angular response (i.e., power pattern), which sets the relative weighting between sky regions of high and low delay, and thus, apparent source frequency structure. We measure the angular response at 137\,MHz using the ORBCOMM beam mapping system of \citet{neben15}. We measure a collecting area of 93\,m$^2$ in the optimal dish/feed configuration, implying HERA-320 should detect the EOR power spectrum at $z\sim9$ with a signal-to-noise ratio of 12.7 using a foreground avoidance approach with a single season of observations, and 74.3 using a foreground subtraction approach. Lastly we study the impact of these beam measurements on the distribution of foregrounds in Fourier space.
\end{abstract}


\section{Introduction}

A new generation of low frequency radio telescopes is coming online with the goal of
 probing redshifted 21\,cm emission from the Cosmic Dawn. These observations will 
 complement indirect probes of the Epoch of Reionization such as quasar 
 sightlines and the CMB optical depth, which leave the reionization 
 history of the universe only loosely constrained. (See \citet{FurlanettoReview, miguelreview, PritchardLoebReview, aviBook, zaroubi} for reviews) In the longer term, 21\,cm observations are expected to improve constraints on cosmology \citep[e.g.,][]{mao08, liu15a,liu15b}. Sensitivity and foreground removal are 
 the main challenges in 21\,cm observations, as the expected cosmological signal is 4--5 
 orders of magnitude fainter in brightness temperature than Galactic and extragalactic foregrounds. Radio 
 interferometers such as the Murchison Widefield Array (MWA) \citep{lonsdale09,tingay13,mwascience}, the Precision Array for Probing the Epoch of Reionization (PAPER) \citep{parsons10,parsons14,ali2015}, the Giant Meterwave Radio Telescope (GMRT) 
 \citep{Paciga2011}, and the Low Frequency Array (LOFAR) \citep{lofar} are seeking a first detection of 
 cosmological 21\,cm emission in power spectrum measurements. In the power spectrum, the spectrally smooth foreground emission separates from the spectrally 
 rough cosmological signal whose frequency dimension probes a line of sight through the 
 inhomogenous reionizing universe.

The Hydrogen Epoch of Reionization Array (HERA) \citep{PoberNextGen,deboer16} is drawing on lessons learned by the MWA and PAPER to reach the calibration and foreground isolation accuracy required to make a significant detection and characterization of the cosmological signal. HERA uses 14\,m diameter parabolic dishes arranged in a compact, hexagonal array to achieve coherent integration of the very low surface brightness 21\,cm signal. Redundant baselines also permit redundant calibration techniques which solve for the relative calibration between all antennas \citep{redundant3, redundant4, liu2010,zheng14}. A central lesson from first generation instruments is that it is essential to characterize the instrument response to foreground emission lest instrument frequency dependence smear foreground power into cosmological signal modes. 


In an ideal achromatic instrument the foreground emission would be confined to the lowest few 
line of sight Fourier modes \citep[e.g.,][]{MoralesBowmanHewittFGsub}, however  the  
interferometer's frequency-dependent point spread function smears foreground power into a ``wedge'' shaped 
region in $(k_\perp,k_\parallel)$ Fourier space \citep{Dattapowerspec,X13, PoberWedge,MoralesPSShapes, VedanthamWedge, nithya13, CathWedge, AdrianWedge1, AdrianWedge2,parsons12b}, where $k_\parallel$ modes are along the line of sight and $k_\perp$ modes are perpendicular to it. This effect is straightforward to understand for a single baseline which measures the sky intensity weighted by the complex sky fringe 
$e^{2\pi i \nu \tau_g}$, where $\tau_g=\vec{b}\cdot\hat{s}/c$ is the delay in radiation arrival time at the second antenna relative to the first antenna of the baseline. Here $\nu$ is the observation frequency, $\vec{b}$ is the baseline vector, and $\hat{s}$ is the direction of the source. 
 Thus sources at different positions relative to the baseline vector appear with different 
frequency structure despite their intrinsically smooth spectra. However, this instrumental frequency structure is limited 
by the baseline length to a maximum frequency dependence of  $e^{2\pi i \nu b/c}$ for sources at maximum delay, near the horizon in line with the baseline vector. This limits the foreground contamination 
to a wedge shaped region in Fourier space with $k_\parallel<a k_\perp$, where $k_\perp$ and $k_\parallel$ represent spatial modes 
perpendicular and parallel to the line of sight, and $a$ is a constant depending on the observational frequency and cosmology. The complement of the wedge is known as the ``EOR window''.

So because sources acquire frequency dependence based on their position on the sky, and the primary beam weights different regions of the sky differently, we see that the primary beam (i.e., the antenna angular response) strongly affects the aggregate frequency dependence 
of the foregrounds. \citet{nithya15} simulate the foreground contamination seen with a dipole beam, a phased array, and a Airy pattern,
and find that the latter suffers the least foreground leakage into
 $k_\parallel>0$ modes due to its narrow main lobe and minimal sidelobe 
levels. To be sure, all are subject to the same geometric limits on foreground frequency-
dependence which limit foreground bounding foreground emission within the wedge, but the emission from high delay is better suppressed using the 
Airy pattern leaving much of the wedge effectively empty. 


For foreground avoidance-based power spectrum estimation, so long as foreground emission is perfectly contained in the wedge it is irrelevant how much or 
little of it there is, but real world effects smear power beyond the geometrical edge of the wedge into the EOR 
window. Finite bandwidth, imperfect bandpass calibration, and faraday rotation of polarized sources can all imprint slight frequency structure on otherwise spectrally smooth sources \citep{jelic2010,giannisurvey, moore2013,moore2015,asad2015,newburgh14,shaw15}, and those closest to the edge of the wedge are 
most at risk of leaking into the EOR window. In fact, 
\citet{nithya15,nithya15b} observe in simulations and then in data that while naively we might expect minimal emission 
at the very edge of the wedge because typical near-horizon beam responses are so small, 
two effects can cause a relative brightening of emission at those maximal delays, creating a characteristic ``pitchfork'' shape. This horizon 
brightening is caused by the large solid angle subtended by the near-horizon regions of the 
sky, as well as the apparent shortening of baselines when viewed nearly on axis at these elevations. 
This second effect makes intermediate length baselines of tens to hundreds of meters sensitive to the very bright diffuse 
emission they would not see from near zenith. Together, these effects can overcome the decline in beam sensitivity near the horizon. 
All these considerations highlight the antenna beam as a critical design parameter for 
21\,cm observatories.


This is the first in a series of four papers detailing the HERA element. In this work we study 
the angular response of the dish and its implications for power spectrum measurements. The three companion 
papers present reflectometry measurements \citep{patra16} and simulations \citep{ewallwice16} of the dish frequency response, as well as detailed 
foreground simulations for HERA \citep{nithya16}. A general description of the design of the 
HERA experiment is given by \citet{deboer16}. In essence, we 
require a large collecting area for
 sensitivity, and minimal sidelobes and horizon response without incurring the large cost per collecting 
area of very large dishes. A dish is preferred to a large phased array as it has a less complex beampattern and reduced potential for antenna-to-antenna variation \citep{neben16}. The core array consists of 320 dishes positioned on a compact, hexagonal 
grid \citep{dillonparsons16} permitting redundant baseline calibration and coherent integration in $
\vec{k}$ space \citep{zheng14,parsons12a}. Improved imaging is permitted by 30 outriggers, but these do not appreciably affect power spectrum sensitivity.

In this paper we first characterize the angular response of a prototype HERA dish at the National Radio 
Astronomy Observatory--Green Bank. We use the beam mapping system of \citet{neben15} to 
measure the 137\,MHz beam pattern using the ORBCOMM satellite constellation. We obtain beam 
measurements out to zenith angles of $\sim60^\circ$ where the beam response is -35\,dB relative to zenith, and compare with different numerical electromagnetic models. We characterize the dish beam at various feed heights to map out the focus and study beam errors due to feed misalignment. We compute the collecting areas and implied EOR power spectrum sensitivities of our measured beams. After verifying our models, we consider the science implications of these 
beam patterns by foreground delay spectra at different baseline lengths and observing conditions to study when the horizon brightening effect is strongest, and thus, when foregrounds are most at risk of leaking into the EOR window.

We discuss the electromagnetic design and modeling of the dish in Section 2. We present the 
experimental setup of the beam mapping experiments and discuss their systematics, then 
review the ORBCOMM beam measurement system in Section 3. We present our power pattern 
measurements in Section 4, and study the science implications of these beam measurements for foreground power spectra in Section 5, then conclude with discussion in Section 6.

\section{Dish Design and Modeling}

\subsection{Design of the HERA Dish}

\begin{figure}[h]
\includegraphics[width=3.4in]{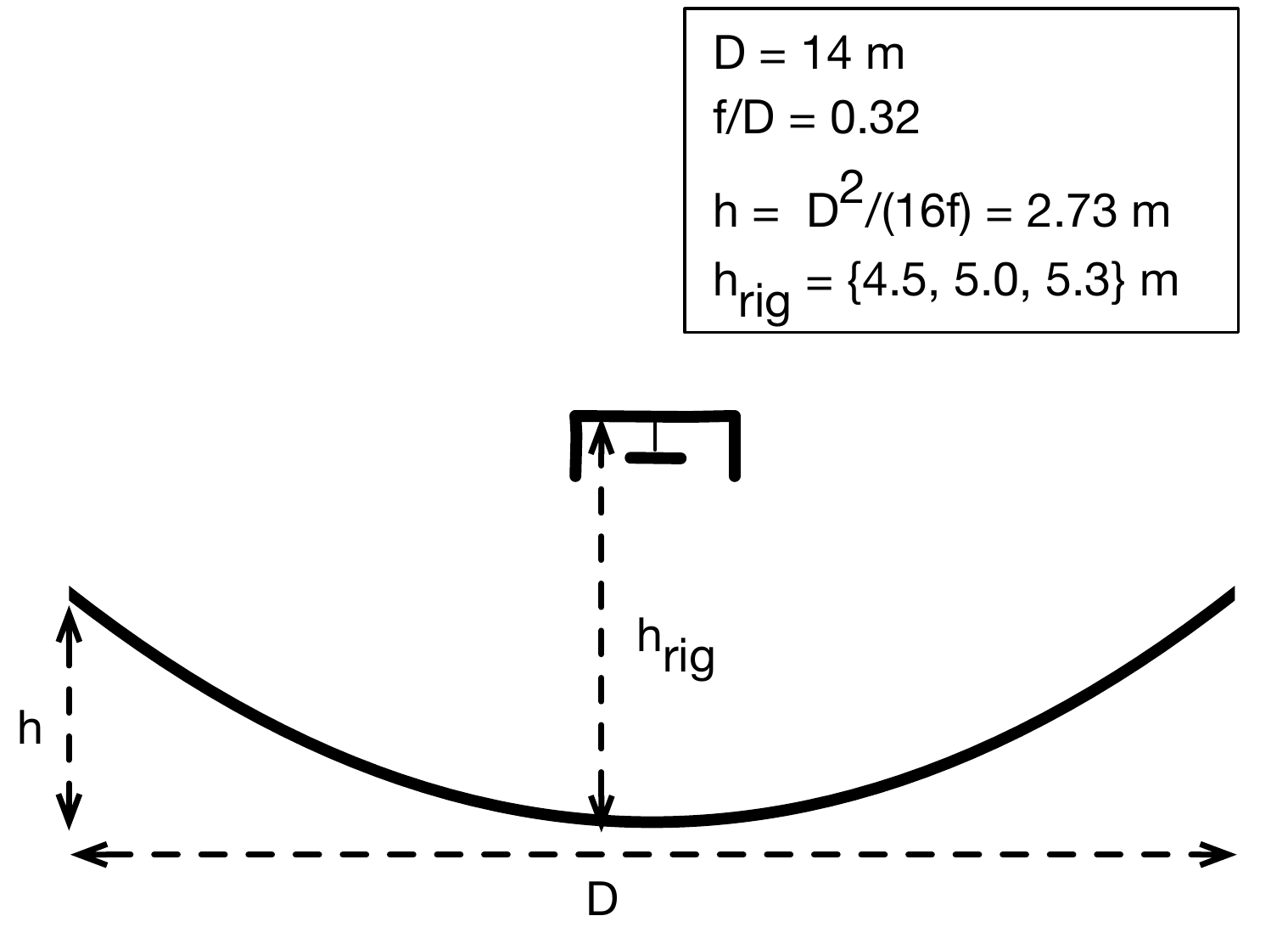}
\caption{Diagram showing the dimensions and layout of the parabolic HERA dish and suspended feed.}
\label{fig:feeddiagram}
\end{figure}

\begin{figure}[h]
\includegraphics[width=3.4in]{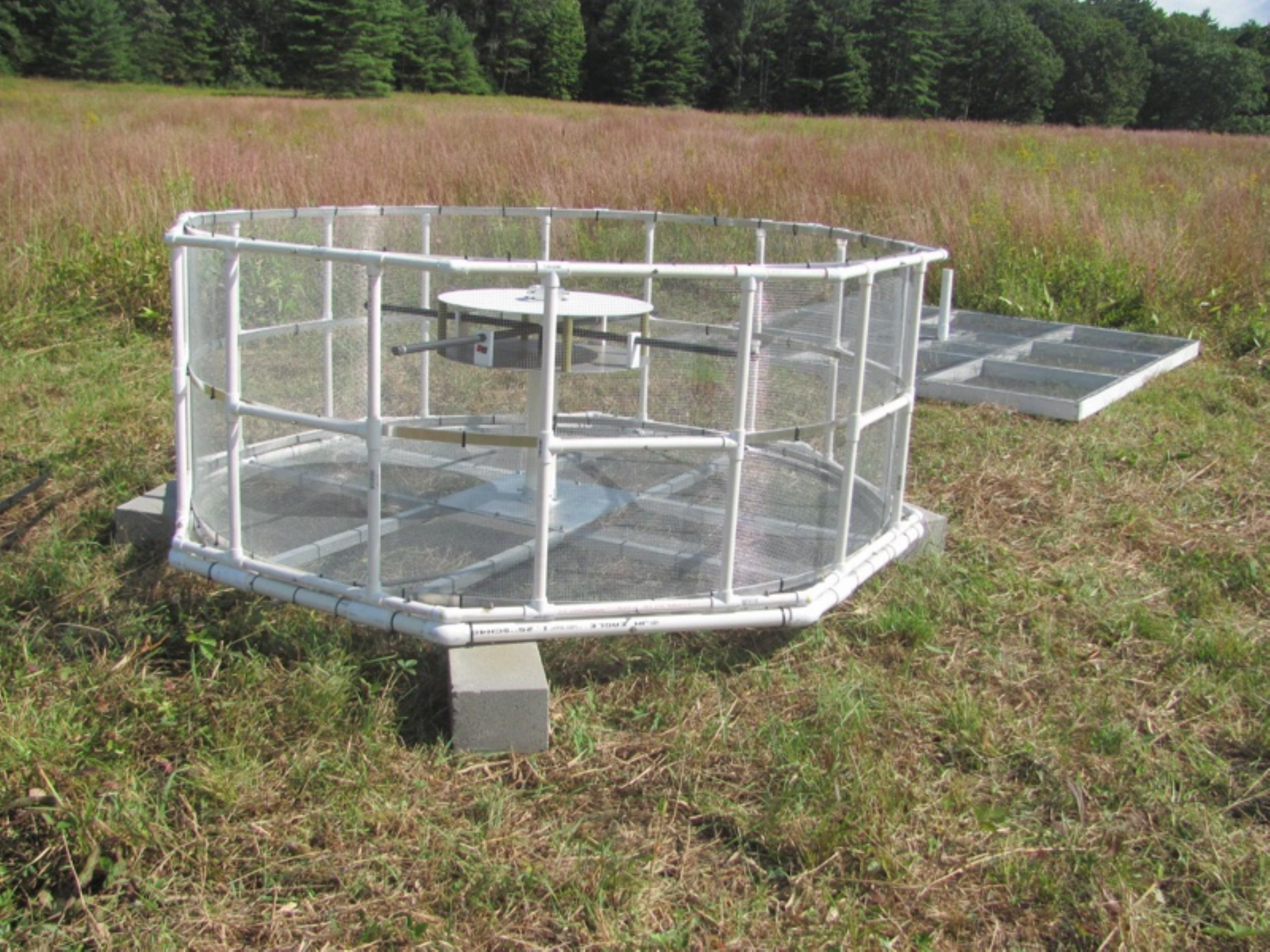}
\caption{Prototype HERA feed seen here outside the dish and upside-down for preliminary characterization. This feed revision consists of a dual-polarized sleeved dipole offset 17'' from a 78'' diameter back plane, surrounded by a 30'' deep cylindrical skirt.}
\label{fig:feedphoto}
\end{figure}

The HERA element (Fig. \ref{fig:feeddiagram}) is a 14\,m diameter faceted paraboloid ($ f/D=0.32$) with a dual-polarized dipole feed suspended at prime focus \citep{dishmemo}. Here $f$ is the focal length of the dish and $D$ is the dish diameter. The dish surface is formed by wire mesh sheets (i.e., facets) mounted on PVC tubes which run from the lip of the dish to the hub at the vertex. For these tests, the feed consists of a dual linear polarization PAPER sleeved dipole mounted 17'' below a 78'' diameter wire mesh back plane surrounded by a 30'' deep cylinder. The feed is suspended from a single point on its back plane from three ropes, each attached to a telephone pole. The three telephone poles are equally spaced around the dish. The dipole ``sleeves'' are circular disks just above and below the dipole designed to broaden its frequency response. The feed cylinder is offset 0.5'' from the back plane, and is designed to make the dipole beam more azimuthally symmetric and also taper its response near the edges of the dish to reduce spillover into adjacent dishes. Fig. \ref{fig:feedphoto} shows the feed as deployed on the ground for early testing. 

The nominal dish focus is $f=(f/D)D=4.48$\,m, though given its faceted design, the dish does not have a single focus. Our numerical electromagnetic models suggest the best focus is slightly higher than that of an perfect paraboloid. In this work we study the dish beam pattern at rigging heights of 4.5\,m, 5.0\,m, and 5.3\,m, measured from dish surface to feed plane, the last height being the maximum height we can achieve with the feed suspension system installed on the dish. These height measurements are uncertain at the $\pm5\%$ level in this study. For more details on the dish design and construction see \citet{deboer16}. Feed optimization studies are ongoing and the values of these parameters may change in the full HERA array \citep{feedoptimizationmemo}. 

As the HERA element is larger than the MWA or PAPER antenna elements, one might worry about the  smaller field of view and thus smaller range of Fourier space probed perpendicular to the line of sight. However, this is a 
small effect for 21\,cm power spectrum analyses as our leverage on $k$ modes comes primarily from modes along the line of sight (in the frequency dimension). Further, HERA's smaller field of view is actually desirable in that it drastically reduces the magnitude of emission at the edge of the wedge compared to a simple dipole element \citep{nithya15}. A second potential drawback is frequency structure introduced by time domain reflections between the dish and feed detailed by Ewall-Wice et al. \citep{ewallwice16} with simulations and \citep{patra16} with zenith reflectometry measurements. These works demonstrate, though, that the slight frequency structure of the dish is sufficiently small to not interfere with EOR science.

\subsection{Dish Modeling}
\label{sec:dishmodels}

\begin{figure*}
\centering
\includegraphics[width=7in]{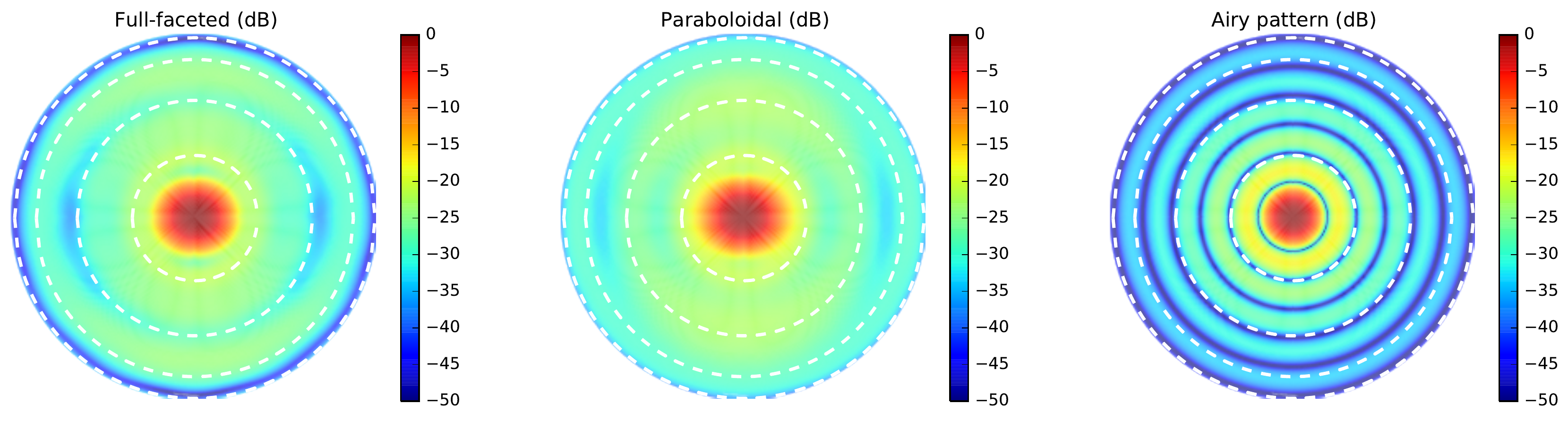}
\caption{Simulated dish power patterns (NS polarization) at 137\,MHz (see Sec. \ref{sec:dishmodels}) with $h_\text{feed}=5$\,m using the full-faceted model (left) and the perfect paraboloidal model (middle) are shown beside an ideal Airy pattern for a 14\,m diameter dish for comparison. Dased lines mark zenith angles of 20$^\circ$, 40$^\circ$, 60$^\circ$, and 80$^\circ$.}
\label{fig:modelbeams}
\end{figure*}

We numerically model the HERA dish in two different ways in order to study the range of realistic beams given modeling inaccuracies and material imperfections. In particular, the near horizon beam response, which sets the level of horizon brightening in the delay spectrum, is quite sensitive to modeling assumptions. We first generate a full-faceted model of the dish using ANSYS HFSS\footnote{http://www.ansys.com/Products/Electronics/ANSYS-HFSS}. All mesh surfaces are modeled as solid aluminum and the dipole itself is modeled as copper. The 1\,m concrete circle at the vertex is modeled with a dielectric similar to dry soil. For comparison, we also model the dish as a perfect paraboloid. We simulate this second model using CST Microwave Studio\footnote{https://www.cst.com/Products/CSTMWS}, but the differences are dominated by the dish geometry, not the choice of numerical electromagnetic solver. 

The simulated full-faceted and perfect paraboloid beams for the NS dipole are plotted in Fig. \ref{fig:modelbeams} (left and center panel) along with an Airy pattern for comparison. As expected, both model beams have slightly stronger sidelobes and wider main lobes than the ideal Airy pattern. The dipole sleeve (circular pieces in Fig. \ref{fig:feedphoto}) and skirt result in a feed beam which is slightly elongated in the E plane and slightly compressed in the H plane, opposite to the behavior of a simple dipole. This wider dish illumination in the NS direction by the NS feed dipole results in a narrower dish beam in the NS direction. Similarly, the EW dish beam is narrower in the EW direction. Lastly, we note that in both models, the best focus is found to be close to 5.23\,m with this feed geometry.

\section{Experimental Setup}

\subsection{ORBCOMM Beam Mapping System Review}
\label{sec:orbcommreview}

We briefly review the beam mapping system detailed by \citet{neben15}, then discuss the 
application of the system for HERA dish measurements. The system 
takes advantage of the 137\,MHz communications satellites operated by ORBCOMM Inc. 
as bright point sources which, by virtue of their number ($\sim30$), short orbital periods 
($\sim90$ minutes), and orbital precession, cover $\sim$65\% of the visible sky in just a few 
days. The coverage from the Green Bank site is limited by the fact that the satellites' orbital inclinations are all less 
than $45^\circ$. 

Unlike celestial source beam measurements, where the flux may be 
assumed constant over the timescale of the measurement, satellite fluxes can vary rapidly 
due to changing distance, orientation, and transmission power. To correct for this, we 
measure the satellite flux in each ground polarization (East-West (EW) and North-South (NS)) using a simple, well-
modeled reference antenna. Comparison of this measured power with that observed in the 
Antenna-Under-Test (AUT) gives the AUT beam response in the direction of the satellite. 
An equivalent interpretation of the measurement is that the power ratio between the AUT and the reference 
antenna gives the relative beam response in the satellite direction, and multiplication by 
the reference antenna model yields the desired AUT response. As discussed in 
\citet{neben15}, this procedure correctly measures the desired response of the AUT to unpolarized radiation despite the fact that satellite signals are generally polarized.

In detail, we measure the dual-polarization RMS power received by each antenna in 512 2\,kHz 
channels across the 137--138\,MHz band. Each band power is averaged over $\sim0.2$
\,sec. There are 0--3 satellites above the horizon at any given time transmitting on different 
$\sim15$\,kHz wide sub-bands in 137--138\,MHz. By observing at many different 
frequencies, we probe the beam response in all these directions simultaneously. We 
compute the satellite positions using the orbital elements published by Celestrak\footnote{http://www.celestrak.com/NORAD/elements/orbcomm.txt} and the orbital 
integrator \texttt{predict}\footnote{http://www.qsl.net/kd2bd/predict.html}. However, the 
satellite frequencies vary occasionally to avoid interference within the constellation. 
\citet{zheng14} use interferometric phases to identify and exclude times when multiple 
satellites are in view. As our data acquisition system makes only total power 
measurements, we instead use an ORBCOMM interface box (typically supplied to 
commercial users of the network) to connect to passing satellites and record their identifier 
and transmission frequency during each pass.

In this way, beam measurements are built up along satellite tracks over the course of 
several days of integration, yielding typically 200--300 satellite passes. Each pass is 
processed separately to identify and exclude times of low signal-to-background when the 
satellite is low in the sky or in the off state of a pulsing sequence. At those times, the 
satellite flux no longer dominates over that of the diffuse Galactic background, and a 
power measurement no longer probes the response in only the satellite direction. The beam 
measurements are then gridded in local Azimuth/Elevation coordinates in HEALPix \citep{healpix} as discussed in Sec. \ref{sec:orbcommreview}.

\subsection{HERA--Green Bank: A three-element prototype array}

A 3-element HERA engineering prototype is being constructed at the National Radio 
Astronomy Observatory--Green Bank. We performed the beam measurements presented in 
this work on the first of these dishes to be constructed, future work will characterize its beam in the presence of the other two dishes once they are constructed. The prototype array is situated in Galford Meadow, approximately 1\,km southwest of the Green Bank Telescope. Note that unlike the full HERA site in the Karoo Desert Radio Astronomy Reserve in 
South Africa, the Green Bank site has trees and foothills, as well as moist ground. Our beam measurements
are sensitive to these effects in addition to the construction imperfections of real world dishes.

\begin{figure}[h]
\includegraphics[width=3.4in]{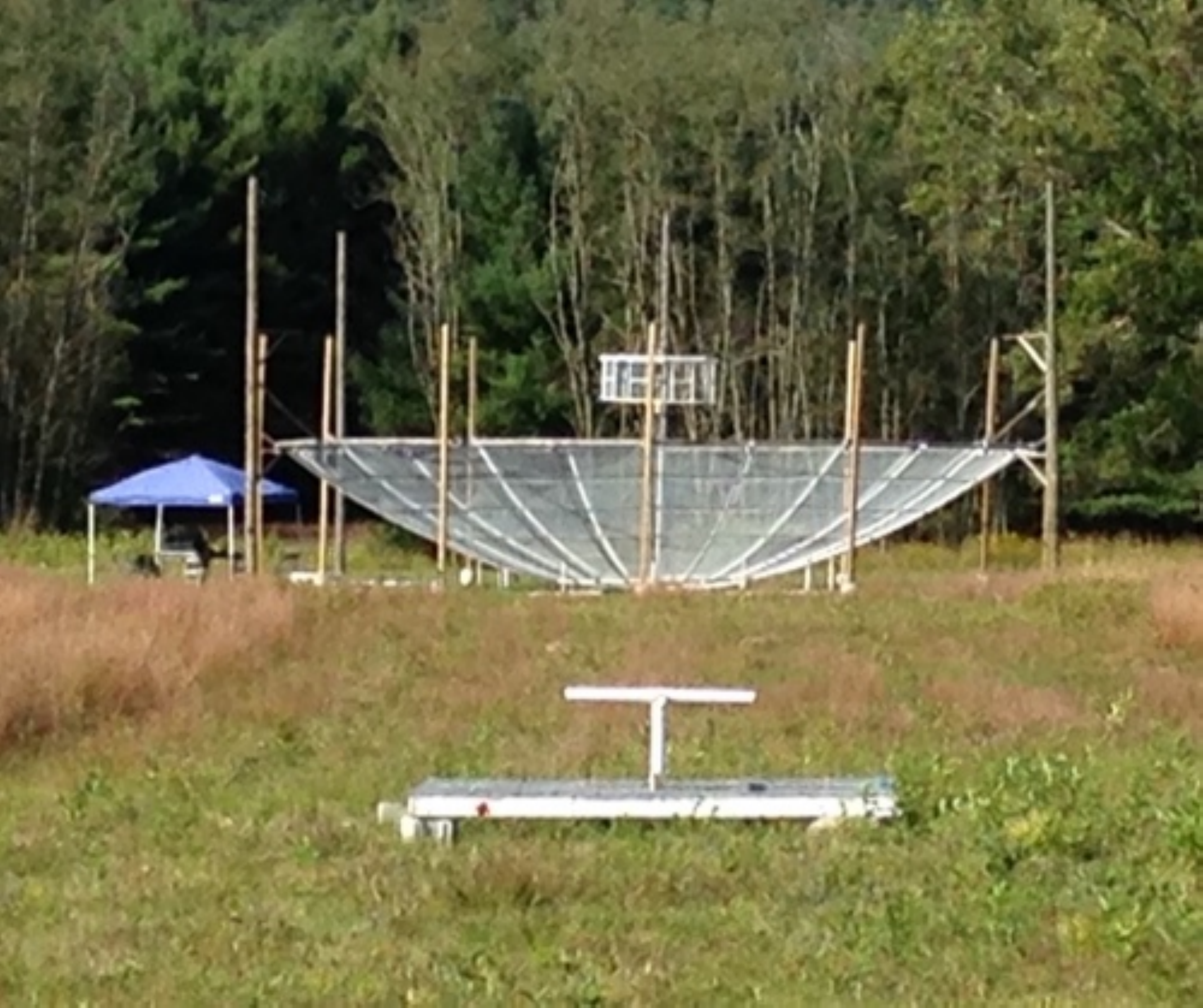}
\caption{The dish with its suspended feed is seen in the back, 50\,m north of one of the reference antennas used in the null experiment to study systematics. The experiment is conducted in Galford Meadow at NRAO--Green Bank.}
\label{fig:greenbankdishphoto}
\end{figure}

We use a simple dual-polarization dipole as our reference antenna. The dipole is constructed out of copper tubing covered by PVC for protection, mounted above a 2\,m $\times$ 2\,m ground plane. See \citet{neben15} for details. During the dish measurements the dipole is positioned 100\,m due south of the dish, though we experiment with other locations in order to characterize the environmental systematics of these measurements, as detailed in the next section. Figure \ref{fig:greenbankdishphoto} shows the dish with suspended feed 50\,m north of one of the reference antennas.

\subsection{Assessing Experimental Systematics}

\begin{figure*}[h]
\centering
\includegraphics[width=6.5in]{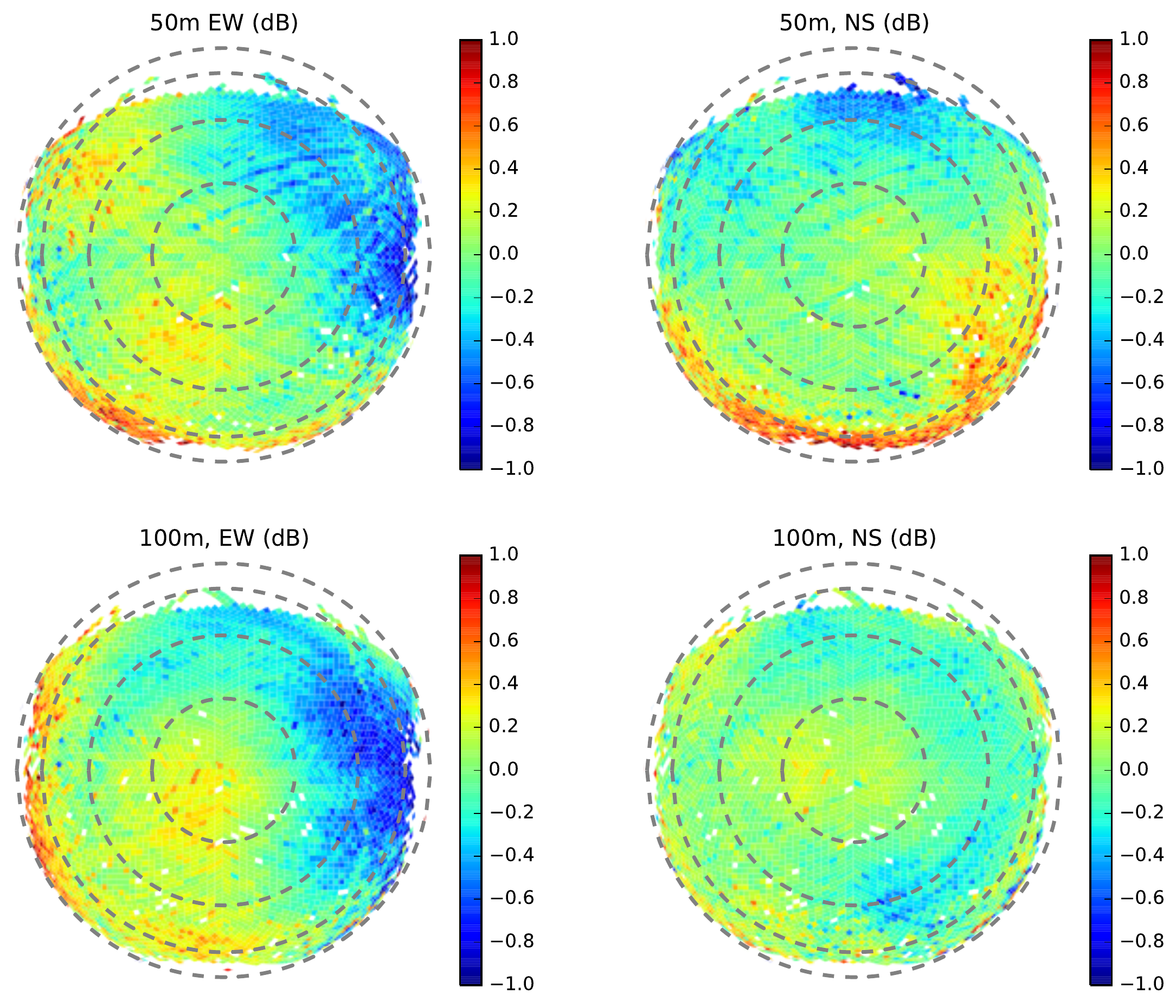}
\caption{We characterize the accuracy of the beam measurement system through null experiments in which a second reference antenna is taken as the AUT and ratio of both reference antenna power patterns is measured for EW (left) and NS (right) polarizations. The reference antennas are separated by 50\,m from each other and from the HERA dish in the first experiment (top), and by 100\,m from each other and from the HERA dish in the second experiment (bottom).}
\label{fig:nullexptplots}
\end{figure*}

As in \citet{neben15}, we assess systematics using a ``null experiment'' in which we use a second reference dipole as the antenna-under-test (AUT). Taking the ratio of its measured power pattern with the model beam pattern amounts to a ratio of the raw power responses received by the two antennas as a function of satellite direction. This probes the level of environmental systematics (i.e., reflections and varying ground properties) and antenna fabrication imperfections which affect each antenna differently. This is not a probe of modeling imperfections common to both antennas, but we expect such errors to be subdominant as the physical properties of the antenna are easier to characterize, and thus simulate, than misalignments and local environmental effects. 

As we are not able to replace the HERA dish with a reference antenna, we run two null experiments with both reference dipoles deployed (1) 50\,m apart on a NS line, 50\,m south of the HERA dish; and (2) 100\,m apart on a NS line, 100\,m south of the HERA dish. Figure \ref{fig:nullexptplots} shows the results from these experiments in the form of the ratio of the power responses of the two antennas. We collected roughly 100 satellite passes. Systematics at the few percent level are observed within $20^\circ$ of zenith, and at the 10--20\% level farther out. The magnitude and angular distribution of these systematics changes modestly as the separation is changed, suggesting that the reference dipoles differ largely due to intrinsic differences, with some environmental variation. In any case, these fractional errors propagate directly into our measured dish power patterns.

\section{Dish Measurements}

\subsection{Power pattern measurements}
\label{sec:powerpatternmeasurements}

\begin{figure*}[t]
\centering
\includegraphics[width=6.5in]{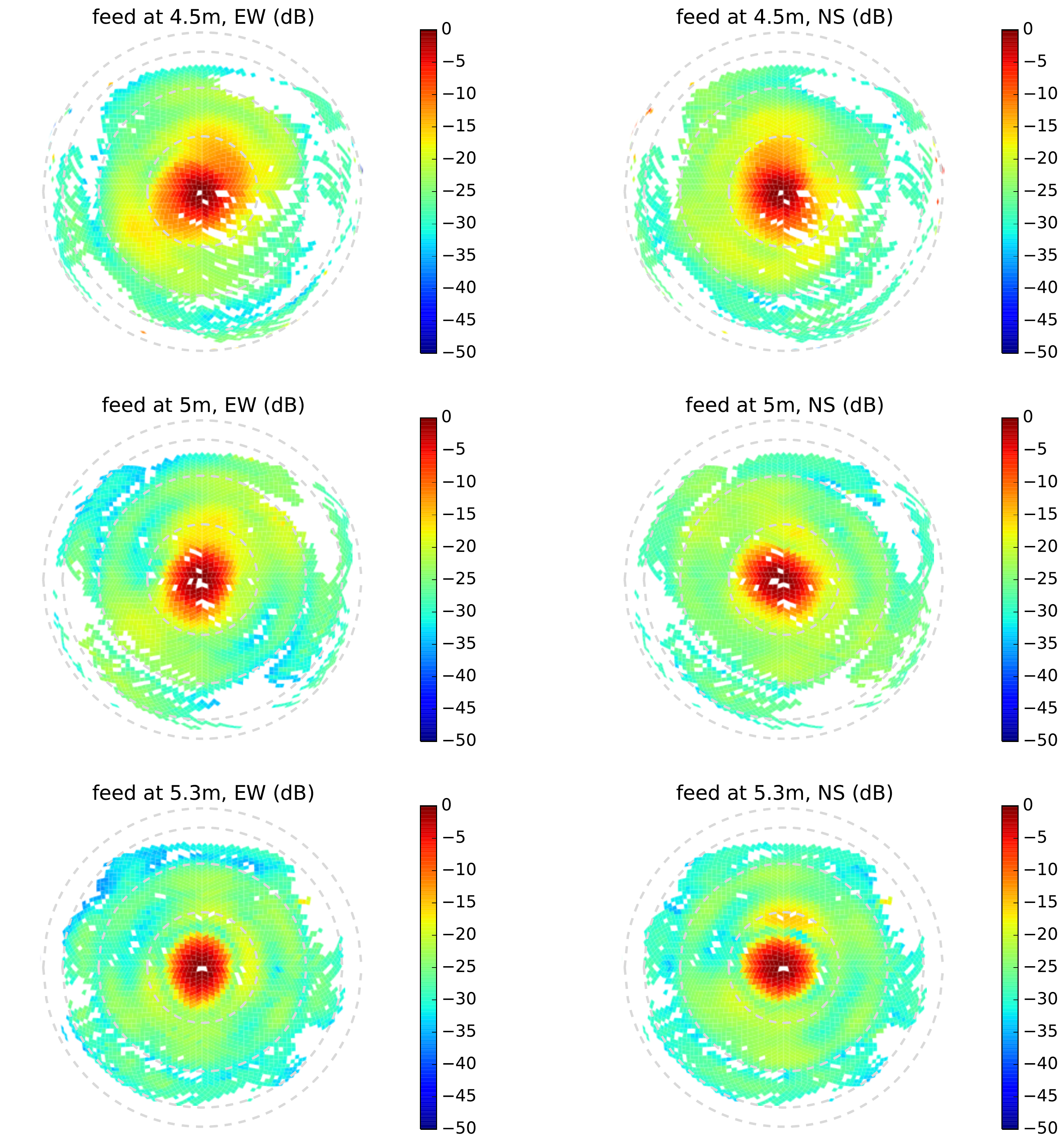}
\caption{Measured dish power patterns at three feed rigging heights (Fig. \ref{fig:feeddiagram}) for the EW (left panel) and NS (right panel) instrumental polarizations. The sidelobes shrink and the main lobe narrows as the feed is raised, confirming that the best focus is close to $h_\text{rig}=5.3$\,m.}
\label{fig:measuredbeammaps}
\end{figure*}

\begin{figure*}[t]
\centering
\includegraphics[width=6.5in]{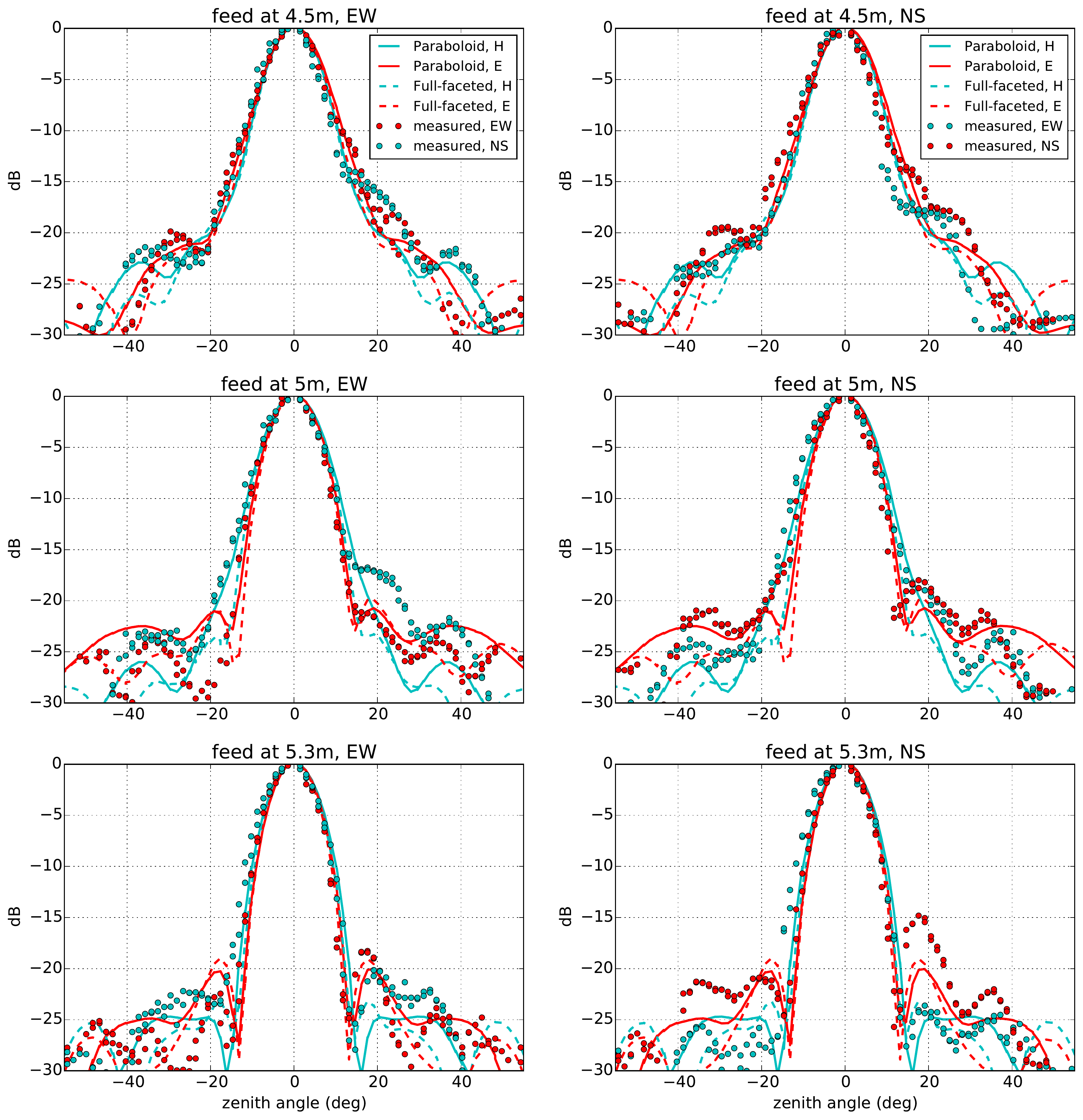}
\caption{Slices through the E (red) and H (cyan) planes through the measured dish power patterns (points) and numerical models (curves). The measured beams agree with both models in the main lobe out to zenith angles of 15--20$^\circ$ up to slight main lobe tilts, but begin to deviate in the sidelobes where the beam response is 25-30\,dB down from zenith. The measured beams typically differ more from both model beams than the models differ from each other, suggesting that real world effects are more significant that the slightly different assumptions used the two beam models. In particular, the most likely systematic is mis-centering of the feed over the dish (see Sec. \ref{sec:powerpatternmeasurements}).}
\label{fig:measuredbeamslices}
\end{figure*}

We make dish power pattern measurements at 137\,MHz as described in Sec. \ref{sec:orbcommreview} with feed rigging heights of 4.5\,m, 5.0\,m, and 5.3\,m above the dish surface (see Fig. \ref{fig:feeddiagram}). In each configuration we collect data for 2--4 days, obtaining roughly 200 satellite passes. We exclude times when the received power is within 20\,dB of the background level determined between passes, and then grid measured beam values into 1.8$^\circ$ HEALPix cells on the sky, rejecting outliers in the top or bottom 5\% in each cell as a final guard against rare satellite identification problems or ADC saturation issues.

Fig. \ref{fig:measuredbeammaps} shows the measured power patterns for these three feed heights for the EW (left panel) and the NS (right panel) feed polarizations. These maps are plotted in sine-projection with dashed circles marking zenith angles of $20^\circ$, $40^\circ$, $60^\circ$, and $80^\circ$. The sky coverage in these dish measurements extends out to typically zenith angles of $\theta\sim60^\circ$, beyond which the ORBCOMM flux is sufficiently attenuated relative to diffuse galactic emission that a power measurements is no longer a clean probe of the antenna gain in the direction of the satellite. At these largest measurable zenith angles the beam sidelobes are roughly -30\,dB down from the zenith boresight gain, and trending downward. 

The roughly $10^\circ$ full-width-at-half-max main lobe narrows slightly as the feed is raised from 4.5\,m to 5.3\,m, and the sidelobes shrink in size and amplitude, confirming that the best focus is closer to 5.3\,m. As discussed in Sec. \ref{sec:dishmodels}, the dish beam should be narrower in the E plane and wider in the H plane, with an overall 180$^\circ$ symmetry. Indeed, the observed main lobes of the EW (NS) beams are slightly wider in the NS (EW), especially in the 5.3\,m feed height beam as it is most in focus. We observe deviations from this symmetry in the sidelobes, which are very sensitive to slight dish/feed imperfections. 

Figure \ref{fig:measuredbeamslices} shows slices through the E and H planes of these power patterns along with the full-faceted and perfect paraboloid numerical models discussed earlier. As in the previous plot, the EW and NS beams are shown in the left and right panels, while the different feed heights are shown in the different rows. The data agree with both models to within 1\,dB in the main lobe, though in several cases appear slightly shifted so they are not quite centered on zenith. The data diverge further in the sidelobes at zenith angles of $20^\circ$ and larger. Here the evolution of the sidelobes as the feed is raised is again seen starkly, as is the fact that the main lobes are slightly wider along the H planes than along the E planes. We observe that both models agree with the measured beams in the main lobe but deviate from the data in different ways at the 1--5 dB level in the sidelobes. Neither model agrees consistently better with the data, suggesting that real-world imperfections of the HERA dish dominate over the slightly different modeling assumptions. 

We emphasize that the model deviations observed in the measured beams are real in that they are larger than the 0.5\,dB scale systematics observed in the null experiments (Fig. \ref{fig:nullexptplots}). Those experiments bound the impact of environmental reflections and reference dipole mismodeling to the 10\% level or smaller across the whole sky. The observed dish beam asymmetries, model deviations in sidelobes, and slight shifts of the main lobes all suggest feed centering errors. The feed is suspended by three ropes attached from the center of the feed back plane to three telephone poles spaced around the dish, and is raised by pulling all three ropes to a new length. Each time this is done the feed centering is slightly disturbed because all three ropes must be pulled to the exact same length to center the feed. Because all three ropes are attached to the same point on the feed, changing their lengths does not affect feed rotation or tilt. Thus if rotation or tilt errors, or dish surface imperfections, were significant, then the beam errors at different feed heights would look similar. The fact that the observed model deviations change with feed height suggests that feed centering errors are most significant. To mitigate all these feed positioning errors, the feeds in the full HERA array will be tied down to the dish surface at several points.

\subsection{Sensitivity}

We compute the effective collecting areas of these beam patterns by first interpolating over unmeasured cells and smoothly extrapolating the power pattern to the horizon. These operations produce a realistically smooth beam which reaches roughly -30\,dB at the horizon, as suggested by the numerical models. The collecting area $A$ is related to the beam power pattern $B(\theta,\phi)$ as
\begin{equation}
	A=\frac{\lambda^2 B(0,0)}{\int B(\theta,\phi)d\Omega}
\end{equation}

The collecting areas are shown in Table 1 along with the maximal collecting area achieved by the Airy pattern for a 14\,m dish. The measured collecting areas imply aperture efficiencies of 45--60\%. This is in line with expectations given the feed design which tapers the dipole beam towards the edges of the dish to reduce spillover into adjacent dishes. The mesh cylinder hanging from the feed back plane around the dipole also reduces the aperture efficiency slightly in order to make the feed beam more azimuthally symmetric. 

 \begin{table}[h]
 \caption{ \label{table:collectingareatable}Collecting area (m$^2$) of measured 137\,MHz beams and corresponding power spectrum SNR for HERA-320 using either foreground avoidance or foreground subtraction.}
\begin{tabular}{| l | l | l |}
\hline
Beam & $A_\text{eff}$ (m$^2$) & SNR ($\sigma$)\\
&& (avoidance, subtraction)\\
\hline
  Airy pattern & 155 & 18.7, 90.8  \\
    Measured, feed at 5.3\,m & 93.0 & 12.7, 74.3 \\
    Measured, feed at 5\,m & 77.1 &  10.6, 67.9 \\
    Measured, feed at 4.5\,m & 68.5 &  10.0, 63.9 \\ 
  \hline
\end{tabular}
\end{table}

We run 21cmSense\footnote{https://github.com/jpober/21cmSense} to compute the overall SNR of a power spectrum detection with one season (6 hours per night for 180 nights) of HERA-320 data. We use a fiducial Epoch of Reionization model generated with 21cmFast \citep{21cmfast}. This model assumes $\zeta=31.5$ for the ionizing efficiency, $T_\mathrm{vir}=1.5\times10^4$\,K for the minimum virial temperature of halos producing ionizing photons, and $R_\text{mfp}=30$\,Mpc for the mean free path of ionizing photons, and reaches 50\% ionization at $z \sim 9.5$ and complete ionization at $z \sim 7$, and is consistent with current observations \citep[e.g.][]{PoberNextGen}. 

We predict SNRs first for a foreground \textit{avoidance} approach where only modes outside of the wedge plus a buffer of $\Delta k_\parallel=0.15\,\mathrm{h}\,\mathrm{Mpc}^{-1}$ are used. These modes have frequency dependence larger than that of any smooth spectrum source on the sky, and this buffer size is chosen to exclude modes which leak out of the wedge due to beam frequency dependence. Due to imperfect impedance matching at the center of the 100-200\,MHz band, the $z\sim8.5$ band requires a slightly larger buffer, though our chosen buffer effectively avoids the leakage in other bands \citep{ewallwice16}. We also predict SNRs for a foreground \textit{subtraction} approach using all modes whose instrumental frequency dependence is larger than that of a source at the edge of the main lobe. 

The SNRs computed with the measured collecting areas are 10-13 with foreground avoidance compared with 19 for the Airy pattern. With foreground subtraction, the SNR falls from 90 with the Airy pattern to 60-75 with the measured collecting areas. In all cases this reduction is a loss of sensitivity, but a power spectrum detection is still always very significant at the 10$\sigma$ level or better.

\section{Foreground Delay Spectrum Simulations}
\label{sec:foregrounds}

We consider now the effects of the beam power pattern on the apparent frequency dependence of the foregrounds. \citet{nithya16} discuss the apparent frequency dependence of foregrounds in more detail as well as the contribution from the beam frequency dependence. We focus in this section on the uncertainties in these foreground power spectrum simulations due to beam modeling uncertainties, but first discuss these foreground simulations themselves and their dependence on observing conditions. 

We simulate foreground power spectra using different primary beam models at various local sidereal times (LSTs). We use frequency-independent model beams (evaluated at 137\,MHz) to isolate the interferometric foreground frequency dependence. The added frequency dependence of the changing overall gain and beam shape with frequency is addressed by the other papers in this series. Given that our measured dish power patterns agree well with both numerical models (full-faceted and perfect paraboloid) in the main lobe but deviate in the sidelobes, and that these models make somewhat different assumptions about the dish surface, we take them as a representative pair of possible dish models. We use the empirically best feed height of 5.3\,m. We also include the Airy pattern for comparison as in \citet{nithya15}. Beam models with weaker response near the horizon (such as the Airy pattern) downweight sources in this direction of high apparent frequency dependence. This reduces the magnitude of emission near the edge of the EOR window, reducing the risk it leaks inside. We use the per-baseline approach of \citet{parsons12a,parsons12b} by first simulating visibilities measured by specific baselines as a function of frequency, then computing the Fourier transform over frequency (delay transform), and lastly normalizing the result into a cosmological power spectrum following \citet{nithya15}. 

In detail, we simulate visibilities using the Precision Radio Interferometry Simulator\footnote{https://github.com/nithyanandan/PRISim} (PRISim) for each beam model at various LSTs, modeling the sky as the sum of the Global Sky Model \citep{gsm} and the NVSS \citep{nvss} and SUMSS \citep{sumss,sumss2} point source catalogs. We use a frequency spacing of 781\,kHz, sufficient to characterize delays within and just outside of the horizon limits on both baseline lengths considered, 14.6\,m and 43.8\,m. We use a total bandwidth of 100\,MHz (effectively reduced to 50\,MHz after applying the Blackman-Harris window) centered on 150\,MHz. This bandwidth is larger than the 10\,MHz thought to be safe from signal evolution with redshift, but is the bandwidth used in the wide band delay space foreground filter of \citet{paper32,ali2015}.

\begin{figure*}[h]
\centering
\includegraphics[width=6in]{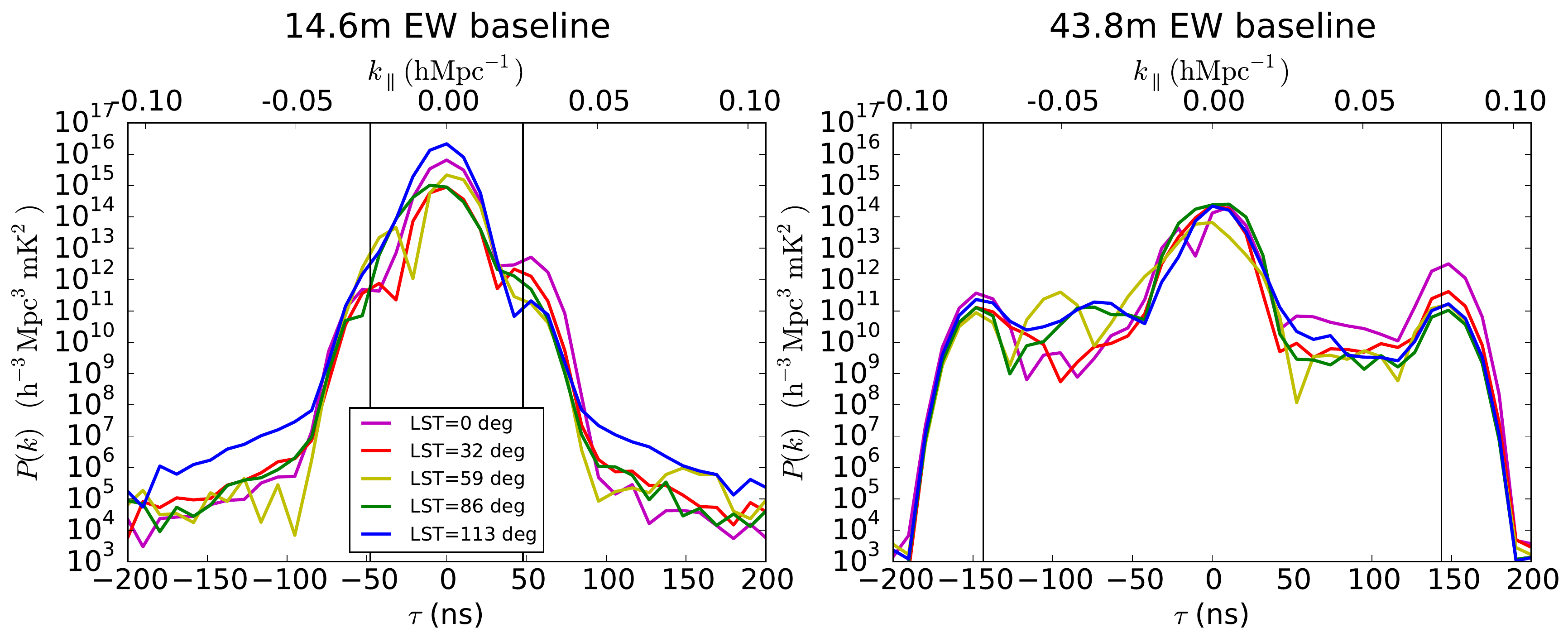}
\includegraphics[width=6in]{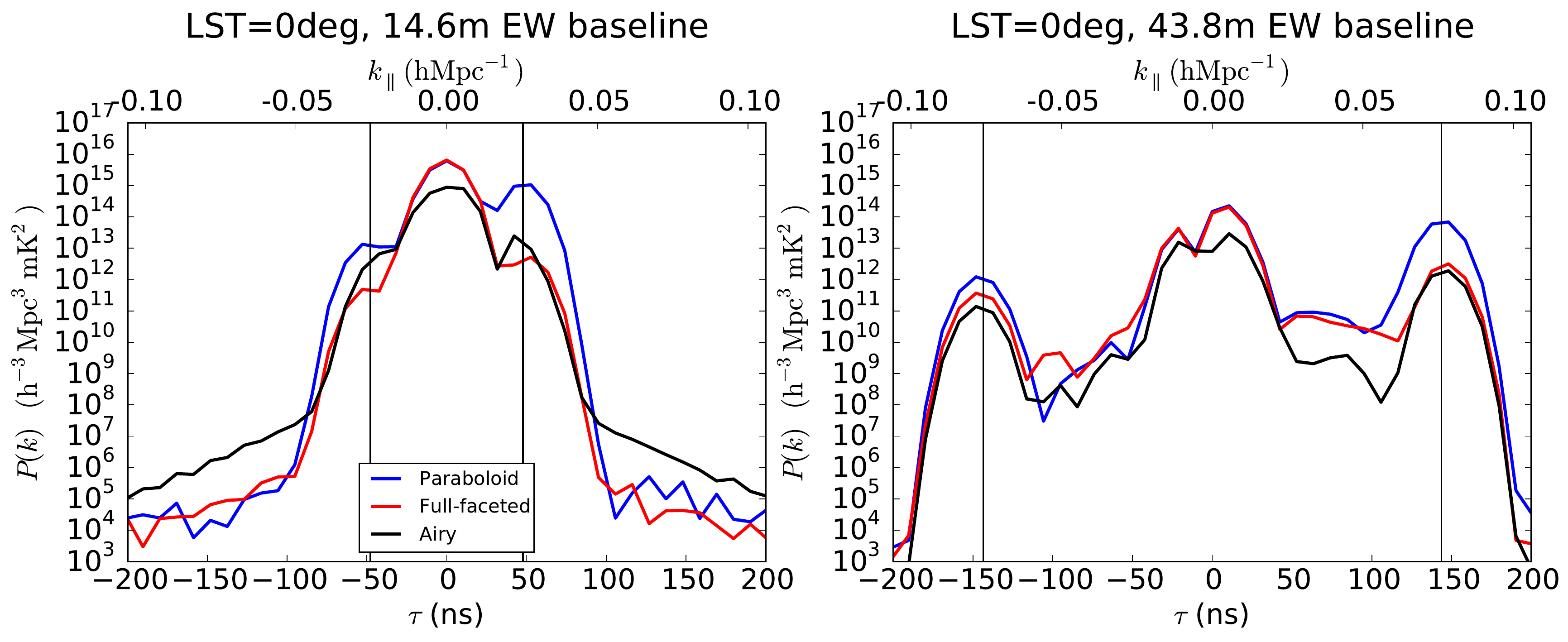}
\includegraphics[width=6in]{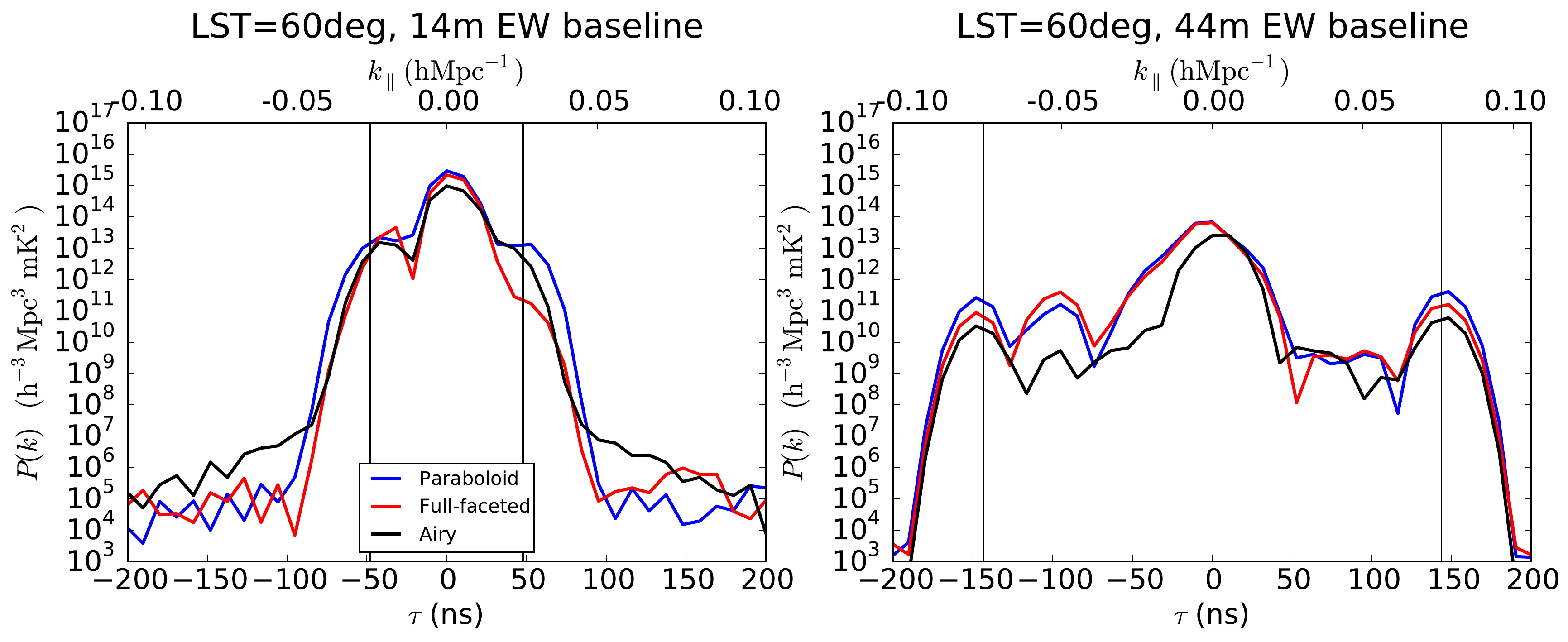}
\caption{We plot simulated foreground delay spectra using the full-faceted beam at various LSTs (top panel). The maximum horizon brightening at the positive horizon occurs close to 0$^\circ$ LST. At this LST, the simulated foreground delay spectra for the three beam models differ markedly near the positive horizon, plotted as a vertical line at the baseline's maximum delay. In contrast, when the horizon brightening effect is smaller at 60$^\circ$ LST (bottom panel), the foreground delay spectra from all three beams agree better.}
\label{fig:delayspec}
\end{figure*}

Figure \ref{fig:delayspec} (top panel) shows simulated foreground delay spectra at various LSTs using the full-faceted beam. As all these LSTs correspond to high galactic latitudes far from the galactic center, the total visibility power (the level of the zero delay mode) varies only by a factor of a few over these LSTs on both baseline lengths (14.6\,m (left panel), 43.8\,m (right panel)). However the positive delay horizon limit (corresponding to the western horizon) has a peak that varies by over 1.5 orders of magnitude on both baselines, demonstrating the stark difference in horizon brightening when the galaxy is just above versus just below the horizon. In this figure we perform the approximate conversion from delay $\tau$ to $k_\parallel$ at $z=8$, which we plot as a second $x$-axis at the top of the plot. 

\begin{figure*}[t]
\centering
\includegraphics[width=6in]{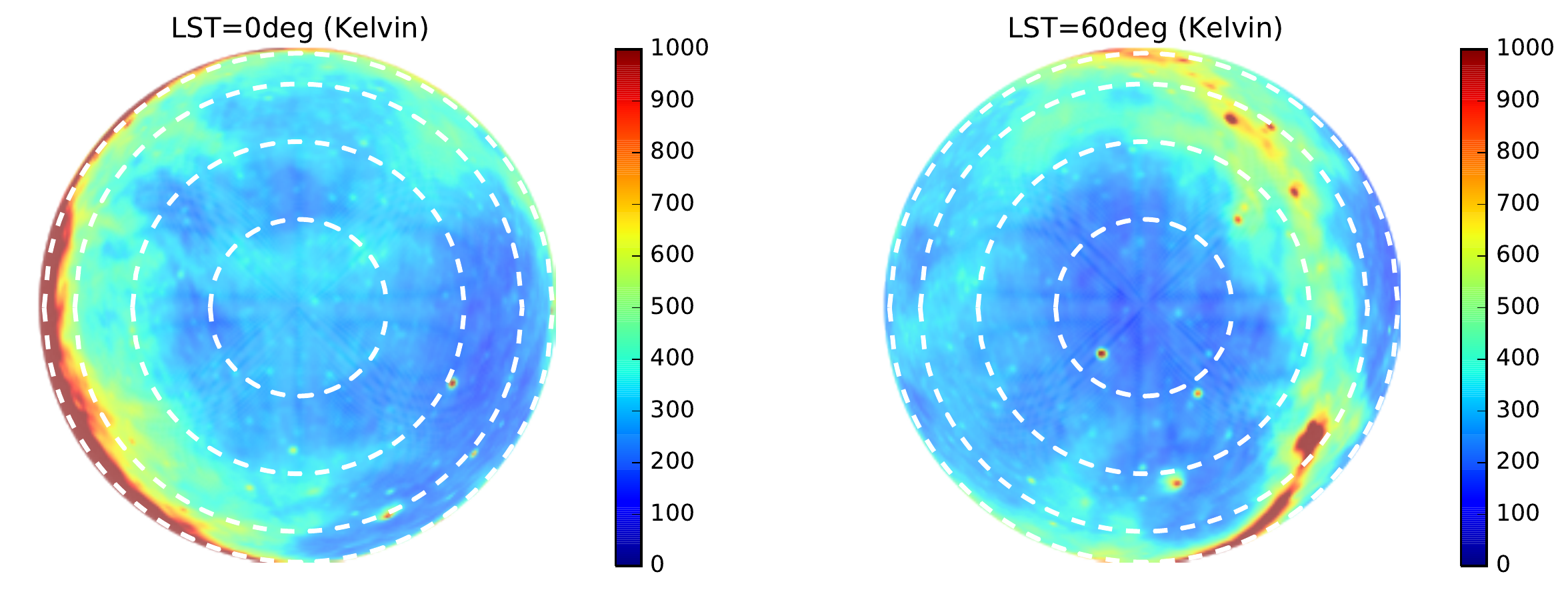}
\caption{Global Sky Model \citep{gsm} in sine-projected horizontal coordinates at LST of 2$^\circ$ (left) and 60$^\circ$ right. The very bright emission from the center of the galaxy at the western horizon at 0$^\circ$ is seen in the delay spectra of EW baselines as a horizon brightening at negative delay.}
\label{fig:gsmplots}
\end{figure*}

To characterize the effect of beam modeling uncertainties on this horizon brightening, we select two of these LSTs, one with maximal horizon brightening ($0^\circ$), and one with minimal horizon brightening ($60^\circ$). Figure \ref{fig:gsmplots} shows the sine-projected Global Sky Model at 150\,MHz, which dominates the horizon brightening effect, in local Azimuth/Elevation coordinates with units of Kelvin for both LSTs. These plots confirm that the large positive delay peak at the 0$^\circ$ LST is due to the center of the galaxy just above the western horizon. In contrast, several hours later, the galactic center is fully below the horizon, leaving only a slight brightening near the eastern horizon due to the weaker galactic anticenter. 

How much do the predicted foreground power spectra differ between the three model dish power patterns? Figure \ref{fig:delayspec} (middle panel) shows the simulated delay spectra for all three beams at $0^\circ$ LST, when the horizon brightening is worst. Both numerical models agree out to delays of roughly 20\,ns on the 14.6\,m baseline and 50\,ns on the 43.8\,m baseline. These numbers suggest that the beams track each other fairly well out to roughly 25$^\circ$ from zenith, beyond which they diverge. This is roughly what is observed in Figure \ref{fig:measuredbeamslices}. At larger delays, especially near the positive delay horizon limit, all three model delay spectra diverge due to the significant edge brightening which effectively discriminates between these models. The perfect paraboloid and full-faceted beams reach roughly -32\,dB and -38\,dB at $80^\circ$ zenith angle (Figure \ref{fig:modelbeams}), consistent with the fact that the perfect paraboloid beam has a larger horizon brightening than the full-faceted beam. This is seen in the delay spectra for both baseline lengths, though the edge brightening is much clearer on the longer baseline where it less diluted by zero delay emission.  

In contrast, all three models agree better when there is little or no edge brightening as in Figure \ref{fig:delayspec} (bottom panel) where we plot the delay spectra for all three beams for 60$^\circ$ LST. There is still a modest flattening off near the horizon on the 14.6\,m baseline and a slight peak on the 43.8\,m baseline due to the large solid angle near the horizon. However as the near horizon emission at this LST is roughly the same temperature as emission from everywhere else on the sky, the difference between the three beam models is greatly reduced.

\section{Discussion}

Power spectrum analyses by first generation 21\,cm observatories are ongoing, but are contending with challenges ranging from calibration and foreground modeling to the analysis effort required to process thousands of hours of data. HERA draws on the most successful ideas from these first generation instruments, pursuing a compact and redundant array layout with large antenna elements. The hexagonal grid allows redundant calibration and coherent power spectrum integration, and the large 14\,m dish achieves sufficient sensitivity at a reasonable data processing and analysis cost. The papers in this series characterize HERA's 14\,m diameter dish element using reflectometry measurements and simulations, which probe its frequency response, as well as power pattern measurements probing its angular response. 

We have presented beam pattern measurements at 137\,MHz, and discussed their implications for 21\,cm power spectrum analyses in terms of sensitivity and foreground isolation. We begun with power pattern measurements made using the beam mapping system of \citet{neben15} which we deployed at the prototype three-element HERA array at the National Radio Astronomy Observatory--Green Bank. We measured the dish power pattern with the feed at different heights over the dish surface and found that the best focus is at a feed rigging height of 5.3\,m, though this may change for different feed designs being explored \citep{feedoptimizationmemo}. The measured beams probe nearly two thirds of the visible sky down to -30\,dB relative to boresight, and agree well with both models in the main lobe out to 10--20$^\circ$ from zenith. The measured beams roughly track the predicted sidelobe levels at 20--30\,dB below zenith, deviating at the 1--5\,dB level.

These deviations away from models and away from 180$^\circ$ azimuthal symmetry are larger than the $\pm0.5$\,dB systematics observed in our null experiments which probe the accuracy of our beam measurement system, suggesting they are genuine measurements of the in situ dish beam. The most likely culprit is feed mis-centering which shifts and distorts the main lobe sidelobes. In the full HERA array, the suspended feeds will be tied to the dish surface at several points to fine tune the feed centering and leveling, and mitigate wind buffeting. Characterizing the level of antenna-to-antenna beam variation in the full HERA array and its effects on power spectrum analyses, as \citet{neben16} do for the MWA, is left as future work.

We quantify HERA's sensitivity to the 21\,cm power spectrum given our beam measurements by first computing the collecting area of the measured beams at the different feed rigging heights, finding 93\,m$^2$ at the best focus, implying an aperture efficiency of 60\%. Feed optimization is ongoing, but the present feed sacrifices aperture efficiency in order to taper the dipole beam towards the edges of the dish and make the X and Y dipole beams as similar as possible using a cylinder hanging from the back plane. We convert our measured collecting areas into effective dish sizes, then use 21cmSense to predict the overall power spectrum SNR at $z\sim9.5$ with one season of HERA-320 data. We predict SNRs of 12.7 and 74.3 using foreground avoidance and subtraction approaches respectively, compared with SNRs of 18.7 and 90.8 using an ideal unobstructed 14\,m aperture (Airy pattern). Still, these sensitivities permit a very significant detection of the 21\,cm signal after a single observing season.

Beyond simple sensitivity considerations, though, the beam pattern affects science analyses by reweighting celestial emission in different regions of the sky, which are then imprinted with different frequency dependence by the interferometer. Longer baselines are more susceptible to this effect, giving rise to a ``wedge'' shaped region in 2D Fourier space. \citet{nithya15} has highlighted that the distribution of foregrounds \textit{within} the wedge is important as well. If the beam falloff is sufficiently shallow at low elevations, there is a relative brightening of emission from near the horizon in line with the baseline due in part to the large solid angle at low elevations. This produces a characteristic ``pitchfork'' shape in the delay spectrum of a single baseline, with a zero delay peak due to bright near-zenith emission surrounded by tines at the negative and positive horizon limits due to emission from the two horizon directions in line with the baseline. These horizon peaks are \textit{most} at risk of leaking foreground power into the EOR window given chromatic instrumental responses such as bandpass miscalibration, though techniques are being developed to suppress emission from near the horizon \citep{parsonsoptimalfringeratefiltering}.

We predict the magnitude of this effect for the HERA element and discuss the uncertainties in this estimate due to beam modeling uncertainty. As expected, we find that the level of horizon brightening is largest when the galaxy is just above the horizon, and lowest when it is well below. When this pitchfork effect is large, we find that the uncertainty in its predicted amplitude is also large, as seen in the differences between the delay spectra calculated using full-faceted and perfect paraboloid beam models. When the effect is small, the two beam models produce much more similar results, highlighting the delay spectrum as an exquisite probe of the difficult-to-measure beam response at very low elevations. Of course the delay spectrum provides only an integrated measure of the beam, but some information can still be extracted. By forward modeling foreground delay spectra using different MWA primary beam models, for instance, it was observed that the MWA bowtie dipoles are better modeled as isotropic radiators than hertzian dipoles at these low elevations (N. Thyagarajan, private communication). Direct measurements using transmitter-equipped drones would be ideal and their development is ongoing \citep{drone1,drone2}.

As discussed by the other papers in this series \citep{ewallwice16,patra16,nithya16}, the frequency dependence of both the beam's angular response and its overall gain widen the delay kernel of a source, leaking power into the EOR window out to $k_\parallel\approx$0.15h\,Mpc$^{-1}$ over much of the 100-200MHz band. This leakage falls within the wedge buffer used in a fiducial foreground avoidance analysis, so our SNR projections take into account the sensitivity reduction due to beam chromaticity. These sensitivities can be improved using new techniques such as foreground covariance downweighting and fringe rate filtering \citep{ali2015,parsonsoptimalfringeratefiltering}, which mitigate foreground leakage into the EOR window, thereby permitting a smaller buffer. Using only these previously demonstrated techniques, we project a $13\sigma$ detection of the EOR power spectrum with a single observing season which would provide begin to probe reionization models in detail and shed light on our cosmic dawn. 

\begin{acknowledgments}
We thank Jonathan Pober, Gianni Bernardi, and Eloy de Lera Acedo for helpful comments on our manuscript. This work was supported by NSF grant AST-1440343, the Marble Astrophysics Fund, and the MIT School of Science. ARP acknowledges support from NSF CAREER award 1352519. AEW acknowledges support from an NSF Graduate Research Fellowship under Grant No. 1122374. DJC acknowledges support from NSF grant AST-1401708.
\end{acknowledgments}




\bibliography{DishBeamMeasurements}

\end{document}